\documentclass[11pt,preprint]{aastex}
\slugcomment{Accepted for the publication in the {\it Astrophysical Journal}}
\def\farcm{\hbox{$.\mkern-4mu^\prime$}}

\def\la{\mathrel{\hbox{\rlap{\hbox{\lower4pt\hbox{$\sim$}}}\hbox{$<$}}}}
\def\ga{\mathrel{\hbox{\rlap{\hbox{\lower4pt\hbox{$\sim$}}}\hbox{$>$}}}}

\shortauthors{Park}
\shorttitle{0540}

\begin{document}

\title{A Deep Chandra Observation of the Oxygen-Rich Supernova Remnant 
0540--69.3 in the Large Magellanic Cloud}

\author{Sangwook Park}
\affil{Department of Astronomy and Astrophysics, Pennsylvania State
University, 525 Davey Laboratory, University Park, PA. 16802, USA}
\email{park@astro.psu.edu}
\author{John P. Hughes}
\affil{Department of Physics and Astronomy, Rutgers University,
136 Frelinghuysen Road, Piscataway, NJ. 08854-8019, USA}
\author{Patrick O. Slane}
\affil{Harvard-Smithsonian Center for Astrophysics, 60 Garden Street,
Cambridge, MA. 02138, USA}
\author{Koji Mori}
\affil{Department of Applied Physics, University of Miyazaki, 1-1 Gakuen 
Kibana-dai Nishi, Miyazaki, 889-2192, Japan}
\and
\author{David N. Burrows}
\affil{Department of Astronomy and Astrophysics,
525 Davey Lab., Pennsylvania State University, University Park, PA. 16802, USA}

\begin{abstract}

Using our deep $\sim$120 ks {\it Chandra} observation, we report on the results from our 
spatially-resolved X-ray spectral analysis of the ``oxygen-rich'' supernova remnant (SNR) 
0540-69.3 in the Large Magellanic Cloud. We conclusively establish the nonthermal nature of 
the ``arcs'' in the east and west boundaries of the SNR, which confirms the cosmic-ray 
electron acceleration in the supernova shock ($B$ $\sim$ 20--140 $\mu$G). We report 
tentative evidence for Fe overabundance in the southern region close to the outer boundary 
of the SNR. While such a detection would be intriguing, the existence of Fe ejecta is not 
conclusive with the current data because of poor photon statistics and limited plasma models. 
If it is verified using deeper X-ray observations and improved plasma models, the presence 
of Fe ejecta, which was produced in the core of the supernova, near the SNR's outer boundary 
would provide an intriguing opportunity to study the explosive nucleosynthesis and the ejecta 
mixing in this young core-collapse SNR. There is no evidence of X-ray counterparts for the 
optical O-rich ejecta in the central regions of the SNR. 

\end{abstract}


\section {\label {sec:intro} INTRODUCTION}

The supernova remnant (SNR) 0540--69.3 in the Large Magellanic Cloud (LMC) is well known 
for its bright central pulsar (PSR B0540--69.3) and its pulsar wind nebula (PWN) which show 
similar characteristics to those of the Crab. The pulsar and PWN have been extensively studied 
at many wavelengths (e.g., Middleditch \& Pennypacker 1985; Manchester et al. 1993a,b; 
Gotthelf \& Wang 2000; Kaaret et al. 2001; Petre et al. 2007 and references therein). SNR 
0540--69.3 is also one of the two known members of ``oxygen-rich'' SNRs in the LMC. In the 
optical band, early data of 0540--69.3 showed a ring-like ($\sim$4$^{\prime\prime}$ in radius) 
[O {\small III}] feature around the pulsar/PWN, which is identified as the O-rich ejecta 
\citep{math80}. The spatial and spectral structures of the optical filamentary features in 
this region were studied using the {\it Hubble Space Telescope} observations, which revealed 
the details of the ejecta and the synchrotron nebula \citep{morse06}. The filamentary features 
of the optical line emission were suggested to result from magnetic Rayleigh-Taylor instabilities 
due to the interaction between the expanding synchrotron nebula and the surrounding ejecta, just 
like those seen in the Crab \citep{morse06}. This central region was also thoroughly studied in 
infrared (IR) using the {\it Spitzer Space Telescope} observations \citep{will08}. These 
observations revealed that the central nebula of 0540--69.3 is analogous to that of the Crab, 
showing synchrotron emission from the PWN, a complex network of ejecta filaments, and newly 
formed dust. A progenitor mass of 20--25 $M_{\odot}$ was suggested based on the observed metal 
ejecta abundances \citep{will08}. The enhanced optical and IR emission most likely originates 
from the interaction between the PWN and the freely-expanding metal-rich ejecta. 

Besides the central pulsar and PWN, the SNR shell produced by the interaction between the blast 
wave and the surrounding medium was detected in the radio band \citep{man93b}. The radio spectral 
index in the outer shell ($\alpha$ = 0.41, where $S_{\nu}$ $\propto$ $\nu^{-\alpha}$) is steeper
than that in the PWN ($\alpha$ = 0.25), which indicated that the shell emission is synchrotron 
radiation from the shock-accelerated electrons. The age of SNR 0540--69.3 has been estimated to 
be $\tau$ $\sim$ 760--1660 yr based on various methods such as the pulsar spin-down, the kinematics 
of the optical ejecta, and the overall dynamics of the ejecta evolutionary models 
\citep{sew84,kir89,rey85,man93a}.    

Thermal emission from the SNR shell ($\sim$1$'$ in diameter) has been detected in the X-ray band 
\citep{hwang01, van01}. The overall SNR morphology of the nearly circular shell with the bright 
W and faint E regions is consistent with that of the radio images. The bright W filaments are 
roughly coincident with the patchy [O {\small III}] emission feature at $\sim$30$^{\prime\prime}$ 
W of the pulsar \citep{math80,sew94}. The detection of these outer filaments, which most likely
originate from interactions between the blast wave and ambient medium, in X-rays and radio makes 
0540--69.3 observationally distinguished from the Crab in which no emission features from the
blast wave significantly beyond the PWN are detected at any wavelengths. Thermal emission from 
the outer SNR shell is detected prominently in X-rays, and thus X-ray observations of 0540--69.3 
provide a unique opportunity to study the ambient and/or circumstellar structures in this SNR 
as well as the reverse shocked metal-rich ejecta. 

Previous X-ray studies of the SNR 0540--69.3 with {\it Chandra} and {\it XMM-Newton} produced 
puzzling results. Enhanced line emission from O, Ne, and Fe was claimed by van der Heyden et al. 
(2001) based on the {\it XMM-Newton} RGS data, but was not detected based on the {\it Chandra} 
observation \citep{hwang01}. The X-ray emission lines detected by {\it XMM-Newton} in the W regions 
of the SNR are strongly blue-shifted ($v$ $\sim$ --2370 km s$^{-1}$, van der Heyden et al. 2001), 
which was not evident in the {\it Chandra} data \citep{hwang01}. The {\it Chandra} data suggested 
an intriguing possibility of the presence of nonthermal X-ray emitting ``arcs'' in the E and W 
boundaries \citep{hwang01}, which are reminiscent of the nonthermal filaments of synchrotron 
radiation from the shock-accelerated cosmic-ray (CR) electrons in several Galactic SNRs such as 
SN 1006. However, the nature (thermal vs. nonthermal) of these faint arcs could not be determined 
with the previous {\it Chandra} data because of the poor photon statistics. The {\it Chandra} 
data also detected a hard tail ($E$ $>$ 3 keV) in the observed X-ray spectrum which appears to 
prevail in the eastern {\it hemisphere} of the SNR \citep{hwang01,hughes01}. The origin for the 
hard tail (a thermal plasma with $kT$ $\sim$ 5 keV vs. power law [PL] continuum with $\Gamma$ 
$\sim$ 2.7) was uncertain \citep{hwang01}. While the previous {\it Chandra} and {\it XMM-Newton} 
data revealed a number of new properties of SNR 0540--69.3, some results were inconclusive and/or 
controversial. The uncertainties and discrepancies in the previous X-ray results appear to be 
caused largely by the poor photon statistics in the {\it Chandra} data (a 28 ks exposure) and/or 
the poor angular resolution ($\sim$1$'$) of {\it XMM-Newton} RGS \citep{hwang01,van01}. Motivated 
by these discrepancies, we performed a deep {\it Chandra} observation to address the outstanding 
issues and to reveal the true nature of SNR 0540--69.3.  

Here we report the results from the data analysis of our deep 120 ks {\it Chandra} observation 
of SNR 0540--69.3. We focus on our spectral analysis of the diffuse X-ray emission features 
outside of the PWN. The spectral analysis of the PWN will be presented elsewhere. In 
Section~\ref{sec:obs}, we describe the observations and the data reduction. We present 
the spectral analysis of the SNR in Section~\ref{sec:analysis}. We discuss the results 
and implications in Section~\ref{sec:disc}, and present a summary in Setion~\ref{sec:sum}.

\section{\label{sec:obs} OBSERVATIONS \& DATA REDUCTION}

We observed SNR 0540--69.3 with the Advanced CCD Imaging Spectrometer (ACIS) on board
{\it Chandra X-Ray Observatory} during AO6 (in 2006 February 15--18, ObsIDs 5549, 7270,
and 7271). We used the ACIS-S3 for the effective detection of the soft X-ray emission
of the SNR. The previous {\it Chandra} observation of 0540--69.3 (ObsID 119) showed 
significant photon pileup effects in the central regions of the PWN \citep{hwang01,petre07}.
To reduce the pileup effects in the PWN as much as possible while covering the entire SNR, 
we chose a 1/4-subarray of the ACIS. We corrected the spatial and spectral degradation
of the ACIS data caused by the charge transfer inefficiency (CTI; Townsley et al. 2000)
using the methods developed by Townsley et al. (2002). We applied the standard data 
screening by status and grade ({\it ASCA} grades 02346). Since our observation was split 
into three segments ($\sim$40 ks exposure for each segment), we performed these data 
reduction steps on individual observations. The three data sets were then merged into 
a single event file by re-projecting them onto the ObsID 5549's tangential plane. After 
the data reduction, the effective exposure was 114.2 ks. The archival {\it Chandra} data 
of 0540--69.3 (ObsID 119) were obtained with a significantly shorter exposure (28 ks), 
and the results of the data analysis have been published \citep{hwang01,petre07}. Since 
the low photon statistics of these archival data did not significantly enhance the statistics
of our new data analysis, we used only our new deep observation in this work. 

\section{\label{sec:analysis} Spectral Analysis}

\begin{figure}[]
\figurenum{1}
\centerline{\includegraphics[angle=0,width=0.9\textwidth]{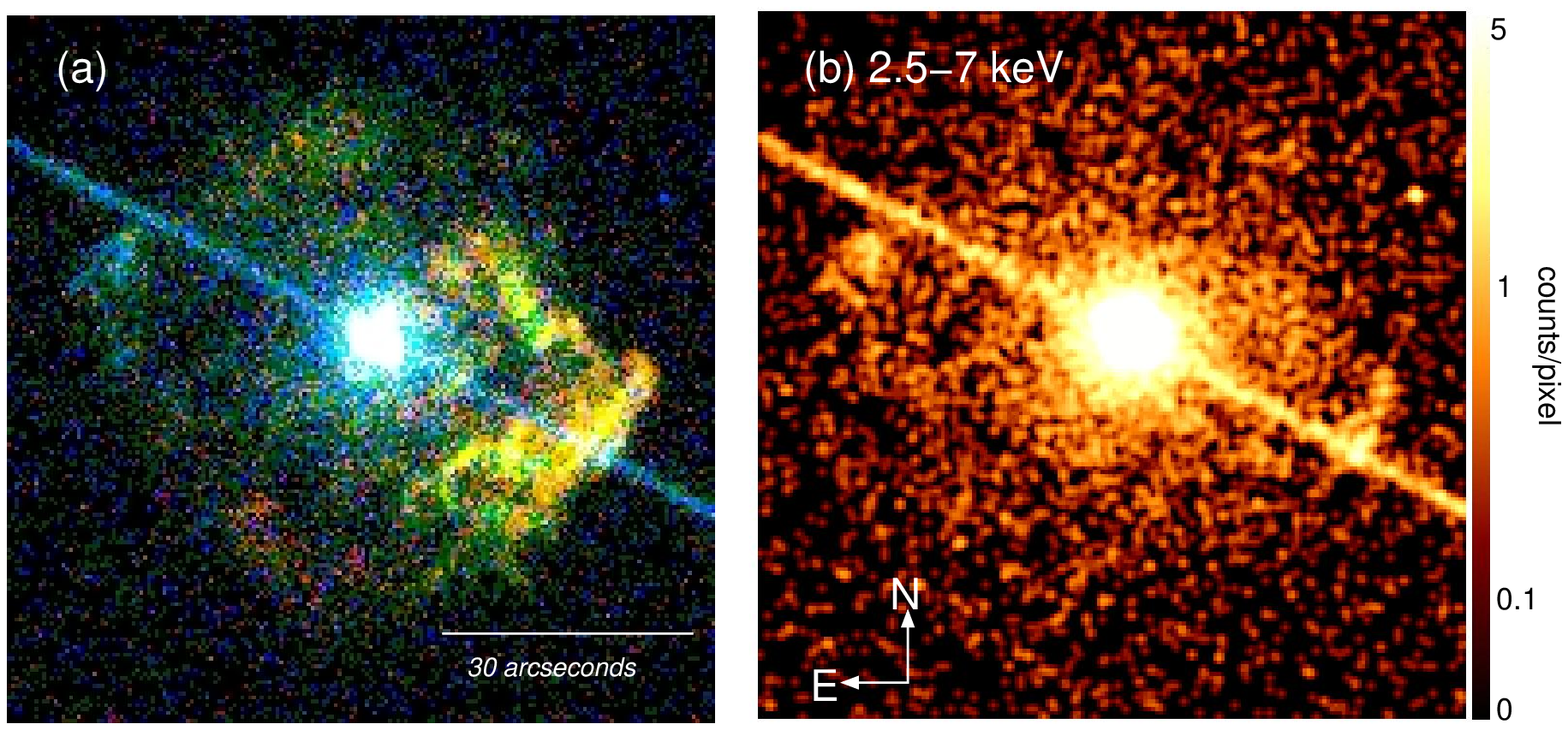}}
\figcaption[]{(a) An X-ray color image of SNR 0540--69.3 taken from the 114 ks
{\it Chandra} observation. Red is 0.4--0.7 keV, green is 0.7--3 keV, and blue 
is 3--7 keV bands. Each subband image is binned into 0$\farcs$492 pixels. 
(b) The hard band (2.5--7 keV) {\it Chandra} image of SNR 0540--69.3. The image
is binned into 0$\farcs$492 pixels. For the purposes of display, the image
is smoothed by a Gaussian with $\sigma$ = 1$^{\prime\prime}$. In both panels,
the central pulsar and PWN are saturated to white to emphasize the faint
filamentary features of the SNR.
\label{fig:fig1}}
\end{figure}

Figure~\ref{fig:fig1}a shows an X-ray color image of SNR 0540--69.3. The central region is 
dominated by emission from the bright pulsar PSR B0540--69.3 and its PWN. The bright X-ray 
emission from the pulsar results in the trail image running NE-SW (corresponding to the
CCD's read-out direction). We excluded this trail region ($\sim$2$\farcs$5 width) in the 
spectral analysis. SNR 0540--69.3 is projected on complex diffuse background structures 
\citep{hwang01}. We tested several nearby source-free regions for the background spectrum. 
Although regions to the S and W of the SNR show generally higher background emission (by 
$\sim$20\% in the total count rates) than regions to the N because of the relatively bright 
soft background emission ($E$ $<$ 1.5 keV) there, the selection of different background 
regions does not have a significant effect on our SNR spectral analysis. In this work, we use 
an average background spectrum extracted from four circular regions ($\sim$15$^{\prime\prime}$ 
in radius) to the NE, NW, SE, and SW of the SNR (Figure~\ref{fig:fig2}). 

Thanks to the use of the 1/4 subarray of the ACIS-S3, our new observation of SNR 0540--69.3
shows reduced photon pileup effects in the PWN region. Based on simple tests using the
event grade distribution and the count rate changes in the SNR between observations in 1999 
and 2006, we find that regions with radial distance $R$ $\ga$ 3$^{\prime\prime}$ from the 
pulsar are nearly free from pileup effects (Regions within $R$ $\la$ 2$^{\prime\prime}$ are
still significantly piled up.) The O-rich ejecta detected in the optical band shows a shell-like
morphology with a radius of $R$ $\sim$ 4$^{\prime\prime}$ \citep{math80,morse06}. Thus, using
the pileup-free data, we can perform a reliable spectral analysis of this region to search
for X-ray counterparts for the optical O-rich ejecta. However, the central regions of the SNR, 
which are spatially coincident with the optical ejecta features, are dominated by the bright 
PWN in X-rays. We find no evidence for enhanced X-ray emission features from the O He$_{\alpha}$ 
and/or O Ly$_{\alpha}$ lines at R $\sim$ 4$^{\prime\prime}$--5$^{\prime\prime}$. This result 
confirms the conclusions inferred by the previous data \citep{petre07}. We here concentrate 
our spectral analysis on the faint E-S regions, bright W filaments, and the faint E arc. 
In all spectral fits presented in the following sections, we binned each regional spectrum
to contain a minimum of 20 counts per energy bin.

\subsection{\label{subsec:arc} The ``Arc'' Region}

\begin{figure}[]
\figurenum{2}
\centerline{\includegraphics[angle=0,width=0.9\textwidth]{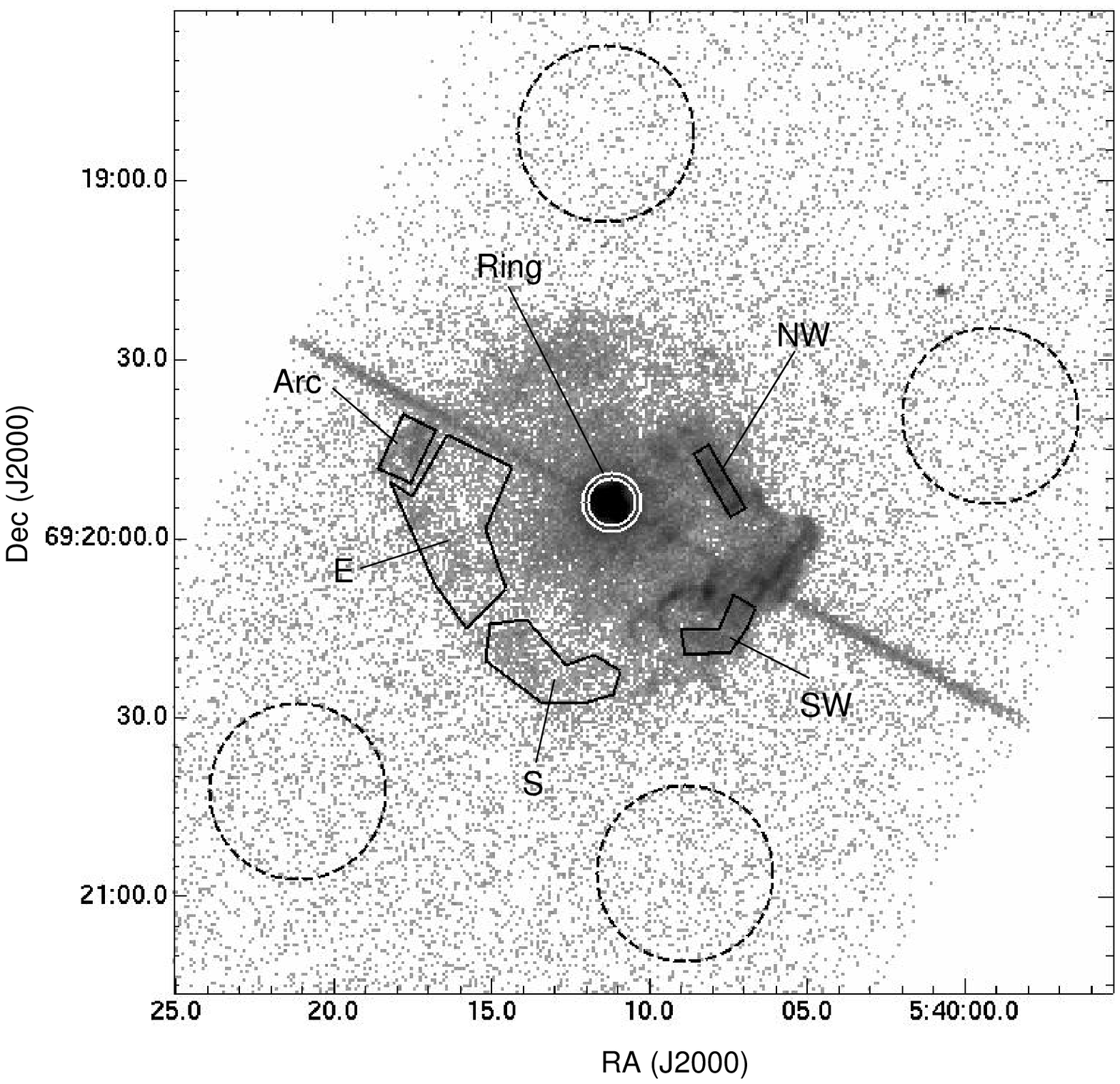}}
\figcaption[]{A grey-scale broadband (0.4--7.0 keV) {\it Chandra} image of SNR 0540--69.3. 
The image is binned into 0$\farcs$492 pixels. Regions of our spectral analysis
are marked. Dashed circles show regions where the background spectrum is estimated. 
\label{fig:fig2}}
\end{figure}

Small arc-like filaments on the E and W boundaries of the SNR were discovered by the previous 
{\it Chandra} data \citep{hwang01}. As Hwang et al. (2001) noted, this particular morphology 
and the hard X-ray spectrum of these features raised the intriguing possibility of strong particle 
acceleration sites at the shock front of 0540--69.3, in analogy to the ``bi-polar'' nonthermal 
X-ray filaments in the Galactic historical SNR 1006. In fact, Hwang et al. (2001) suggested that a 
significant portion of the X-ray emission in these arcs could be nonthermal continuum in origin. 
However, because of the poor photon statistics (e.g., $\sim$190 counts in the arc region in the 
E boundary), the true nature of these features could not be revealed by the previous {\it Chandra} 
data. With a four times deeper exposure, we obtain significantly better photon statistics from 
these arcs to study the spectral characteristics. We note that the W arc region is projected on 
the bright thermal emission features of the SNR. Also, the trail image by the pulsar is partially
superposed onto the W arc (Figure~\ref{fig:fig1}b). The spectral analysis of the W arc is difficult 
and less useful than the E arc due to these complex environments and the detector artifacts. In fact, 
our two-component spectral model fits (plane shock + PL) for the W arc result in nearly the same 
spectral parameters for the hard PL component ($\Gamma$ $\sim$ 2.4, $f_{\rm 0.5-8~keV}$ $\sim$ 4 
$\times$ 10$^{-14}$ erg cm$^{-2}$ s$^{-1}$) as those for the E arc (see below). Because the soft 
thermal emission dominates the overall W arc spectrum, the best-fit PL parameters are poorly 
constrained compared to those in the E arc. Thus, we discuss the nature of the arcs based on the 
spectral analysis of the E arc (the ``Arc'' region in Figure~\ref{fig:fig2}). 

\begin{figure}[]
\figurenum{3}
\centerline{\includegraphics[angle=-90,width=0.9\textwidth]{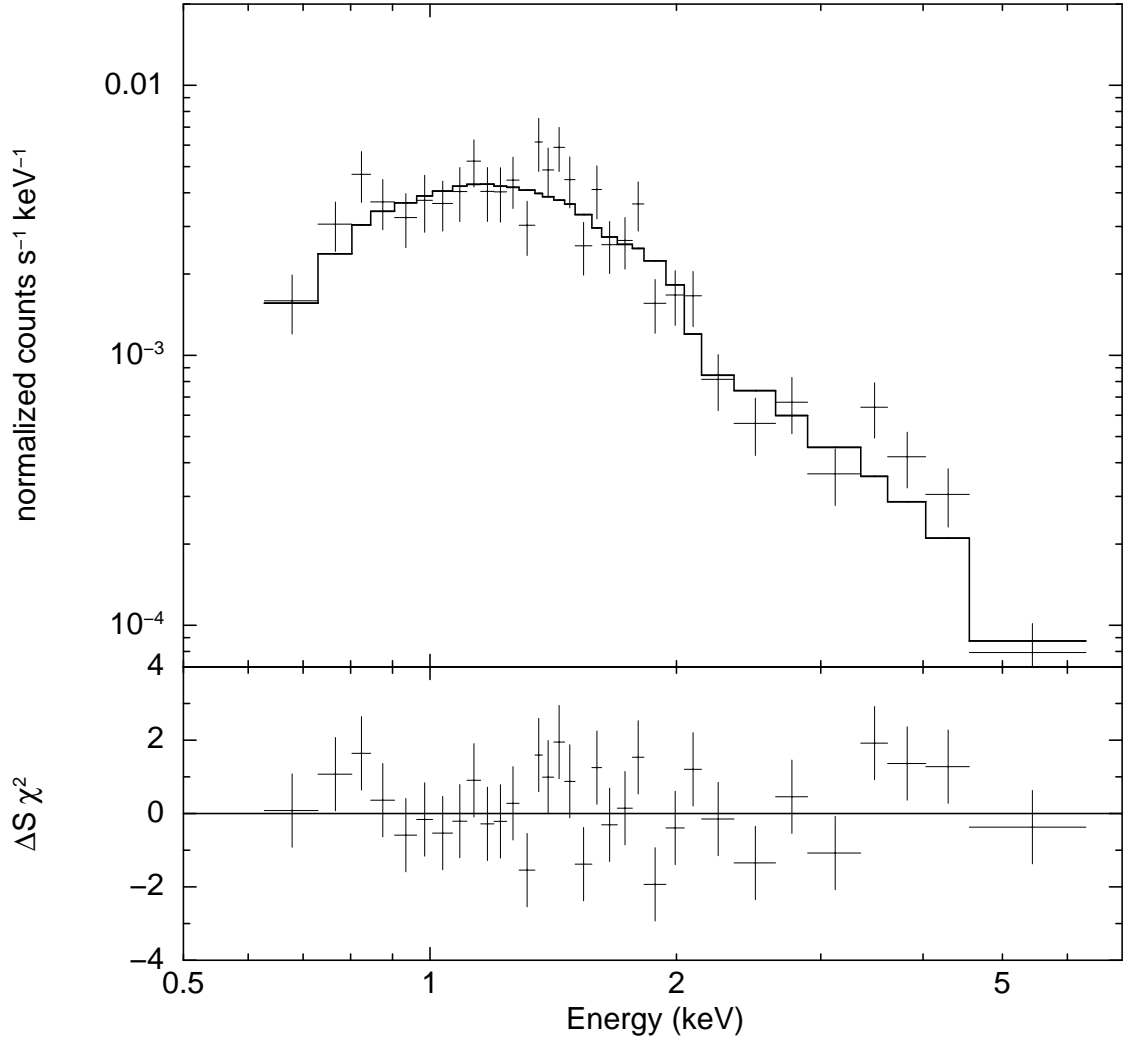}}
\figcaption[]{The {\it Chandra} ACIS spectrum of the Arc region. The best-fit PL
model is overlaid. The lower panel shows the residuals from the best-fit model.
\label{fig:fig3}}
\end{figure}

The Arc region contains $\sim$750 counts and is well-separated from the bright thermal filaments, 
as well as avoiding contamination by the pulsar trail image. The Arc region shows a featureless 
X-ray spectrum which can be fitted by a PL model with $\Gamma$ = 2.42$^{+0.26}_{-0.23}$ 
(uncertainties are with a 90\% confidence level, hereafter). We fixed the Galactic absorbing 
column at $N_{\rm H,Gal}$ = 7 $\times$ 10$^{20}$ cm$^{-2}$ \citep{dl90}. We fit the foreground 
column in the LMC ($N_{\rm H,LMC}$ = 6.5$^{+1.8}_{-1.5}$ $\times$ 10$^{21}$ cm$^{-2}$), assuming 
LMC interstellar abundances \citep{rd92}. The X-ray spectrum of the Arc region and the best-fit 
PL model ($\chi^2/{\nu}$ = 39.8/31) are presented in Figure~\ref{fig:fig3}. The plane-parallel 
shock model with LMC abundances cannot fit the observed spectrum ($\chi^2/{\nu}$ = 2.8). When 
the individual metal abundances are varied, the fit results in a large column ($N_{\rm H,LMC}$ 
$\sim$ 1.9 $\times$ 10$^{22}$ cm$^{-2}$) and a high electron temperature ($kT$ $\sim$ 2.3 keV). 
The fitted abundances are negligible for most species except for O ($\sim$5) and Ca ($\sim$4, 
abundances are with respect to Solar [Anders \& Grevesse 1989], hereafter). The fitted LMC column 
is implausibly large for this location of the LMC. The inferred overabundances for O and Ca are 
statistically insignificant due to large uncertainties. The best-fit model ($\chi^2/{\nu}$ = 
35.0/22) is significantly poorer than the PL model fit. A two-component model (PL + plane shock) 
indicates only a small contribution ($\sim$5\% of the total flux) from the thermal plasma ($kT$ 
$\sim$ 0.4 keV). The PL + plane shock model does not improve the overall fit ($\chi^2/{\nu}$ 
= 36.7/28). Thus, we conclude that the X-ray emission in the Arc region is dominated by nonthermal 
PL continuum. The best-fit photon index ($\Gamma$ $\sim$ 2.4) is consistent with the typical value 
for synchrotron radiation by relativistically accelerated electrons in the strong SN shock.    

Assuming that synchrotron radiation from the shock-accelerated electrons is responsible for
the X-ray and radio emission for the Arc, we fit the observed Arc spectrum with the synchrotron 
cutoff model (SRCUT, Reynolds 1998; Reynolds \& Keohane 1999). We fixed the radio spectral index
at $\alpha$ = 0.41 as estimated for the entire SNR excluding the central PWN \citep{man93b}. 
The best-fit LMC column, $N_{\rm H,LMC}$ = 5.7$^{+1.0}_{-0.9}$ $\times$ 10$^{21}$ cm$^{-2}$, 
is consistent with that measured by the PL model fit. The fitted synchrotron roll-off frequency 
is $\nu_{\rm roll}$ = 1.0$^{+1.4}_{-0.4}$ $\times$ 10$^{17}$ Hz. Our X-ray spectral analysis
indicates that the 1 GHz flux for the radio counterpart is $f_{\rm 1 GHz}$ $\sim$ 0.19 mJy. 
On the other hand, based on the 1.5 and 5 GHz images \citep{man93b}, we roughly estimate 
$f_{\rm 1 GHz}$ $\sim$ 7 mJy for the Arc region. This radio flux is significantly larger than our 
estimate from the X-ray spectral fit, which suggests that the radio spectral index may not be uniform 
across the SNR. Thus, we repeated the SRCUT model fit of the Arc region by fixing $f_{\rm 1 GHz}$ = 
7 mJy while varying the radio spectral index $\alpha$. The best-fit parameters are $N_{\rm H,LMC}$ = 
5.7$^{+1.0}_{-0.8}$ $\times$ 10$^{21}$ cm$^{-2}$, $\nu_{\rm roll}$ = 2.1$^{+4.7}_{-1.2}$ $\times$ 
10$^{17}$ Hz, and $\alpha$ = 0.62$\pm$0.01. The best-fit radio spectral index is not unusual
for Galactic shell-type SNRs \citep{green09}. Based on the PL and SRCUT model fits, the observed 
X-ray flux (absorbed) of the Arc is $f_{\rm 0.5-8~keV}$ $\sim$ 3.7 $\times$ 10$^{-14}$ erg 
cm$^{-2}$ s$^{-1}$. At the distance of $d$ = 50 kpc for the LMC ($d$ = 50 kpc for SNR 0540--69.3 
is assumed in this paper hereafter), the unabsorbed X-ray luminosity is $L_{\rm 0.5-8~keV}$ $\sim$ 
1.7 $\times$ 10$^{34}$ erg s$^{-1}$. The best-fit model parameters for the Arc region are summarized 
in Table~\ref{tbl:tab1}. 

We note that the estimated absorbing column to the Arc can be reliably assumed for other regions 
of SNR 0540--69.3 because it is measured by straightforward models (PL and SRCUT) of the simple 
continuum-dominated spectrum. In fact, our spectral fits for the central region ($R$ = 
4$^{\prime\prime}$--5$^{\prime\prime}$, the ``ring'' region in Figure~\ref{fig:fig2}) also indicate 
$N_{\rm H,LMC}$ = 6.0$\pm$0.5 $\times$ 10$^{21}$ cm$^{-2}$. This central region is close to the PWN, 
in which the observed spectrum is dominated by the simple PL continuum and provides significant 
photon statistics ($\sim$3600 counts), while the pileup effects are convincingly ruled out. We find 
that other regional spectral analysis requires more complicated thermal spectral modelings which 
result in less reliable estimates of $N_{\rm H,LMC}$ (e.g., Hwang et al. 2001). Thus, based on the 
results from the spectral analysis of the Arc region, we fix the LMC column at $N_{\rm H,LMC}$ = 
6 $\times$ 10$^{21}$ cm$^{-2}$ in the following spectral analysis of the SNR. This $N_{\rm H,LMC}$ 
value is somewhat higher than those estimated/adopted in previous works ($N_{\rm H}$ $\sim$ 4--5 
$\times$ 10$^{21}$ cm$^{-2}$, e.g., Kaaret et al. 2001; Petre et al. 2007). The higher $N_{\rm H,LMC}$ 
is justified by the realistic assumption of the LMC abundances in our absorption model, while the 
previous works assumed the Solar abundances. Our $N_{\rm H,LMC}$ is fully consistent with that 
estimated by Serafimovich et al. (2004) who also assumed the LMC abundances for the absorbing column 
by the LMC. 

\subsection{\label{subsec:east} Faint East Region}

The eastern half of the SNR is generally faint, and X-ray emission appears to originate from a 
hot plasma of the electron temperature $kT$ $\sim$ 5 keV \citep{hwang01}. However, the previous 
results on the faint eastern regions were rather preliminary based on a simple spectral analysis 
of the integrated spectra from the ``inner'' and ``outer'' regions of the entire eastern half of 
the SNR. The inner parts of the eastern half of the SNR could be substantially contaminated by 
scattered photons from the bright pulsar and PWN. The outer regions of the eastern half show an 
intensity variation with a relatively bright emission in the north. Furthermore, our new deep 
{\it Chandra} data reveal that X-ray emission near the southern boundary, which was included in 
the ``outer east'' region by Hwang et al. (2001), is spectrally-soft (Figure~\ref{fig:fig1}a). 
Thus, we carefully re-analyze the faint E region excluding the relatively bright N regions and 
the spectrally-soft S regions. We also exclude the central regions of $R$ $\la$ 20$^{\prime\prime}$ 
to avoid the contamination from the pulsar and PWN as much as possible (our radial cut to remove 
the PWN is more conservative than that by Hwang et al. [2001] who excluded regions of $R$ $\leq$ 
7$^{\prime\prime}$.) Based on our MARX\footnote{http://space.mit.edu/CXC/MARX/} simulations of 
the pulsar/PWN of 0540--69.3, we estimated that the contamination by scattered photons from the 
pulsar/PWN due to the broad PSF wings of the mirror is $\sim$20--30 \% in the observed 3--7 keV 
band flux for the ``E'' region (as shown in Figure~\ref{fig:fig2}). In these simulations, we 
assumed our observed ACIS image of the PWN ($R$ $\leq$ 4$^{\prime\prime}$) with a PL spectrum of
$\Gamma$ = 2 and the 1--10 keV band flux of 3.2 $\times$ 10$^{-11}$ erg cm$^{-2}$ s$^{-1}$
\citep{hira02}. The spectrum of the scattered photons is hard (a PL with $\Gamma$ $\sim$ 0.25) 
at these large radial distances of $R$ $\sim$ 20$^{\prime\prime}$--30$^{\prime\prime}$ from 
the pulsar. Thus, this instrumental artifact contaminates primarily at $E$ $\ga$ 5 keV. We fixed 
this scattered photon flux spectrum in the spectral model fits of the E region. (The broad PSF wing 
flux was estimated in the annular region of $R$ = 20$^{\prime\prime}$--30$^{\prime\prime}$ using 
our MARX simulations, and was then normalized for the area of the E region. This normalized flux 
and the best-fit PL photon index for the scattered photon spectrum were fixed in the spectral 
model fits of the E region.)   

The X-ray spectrum of the E region includes $\sim$1600 counts. The overall spectral characteristics 
are consistent with those shown for the ``outer east'' region in Hwang et al. (2001): the soft band 
spectrum evidently shows X-ray line emission features, and the hard tail extends to $E$ $>$ 4 keV. 
Initially, we fit the observed spectrum with a non-equilibrium ionization (NEI) plane-shock model 
(Borkowski et al. 2001; {\tt vpshock} in conjunction with the NEI version 2.0 in the XSPEC) that
is based on ATOMDB \citep{smith01}. We use an augmented version of this atomic database to include 
inner-shell processes and updates of the Fe L-shell lines, whose effects are significant in highly 
under-ionized plasma spectrum, but are not accounted for in the current XSPEC NEI version 
2.0.\footnote{The augmented APEC atomic data have been provided by K. Borkowski. The relevant 
discussion on these detailed plasma model issues has been presented in Badenes et al. (2006).} 
This model includes inner-shell Fe L lines from Na-like Fe ion at $E$ = 725.15, 727.06, and 
738.98 eV which are critical to properly fit the broad Fe L line complex-like feature in the 
0.7--0.8 keV band (see Section 3.3), but were not included in the standard XSPEC NEI version 2.0. 
Thus, although it is still incomplete, our NEI model is the most reliable one that is currently 
available to the astronomical community to fit highly under-ionized plasma. We use this latest 
NEI model in the spectral analysis of thermal X-ray emission throughout this work. We fixed the 
individual metal abundances at the LMC values \citep{rd92}. We fixed the LMC column at $N_{\rm 
H,LMC}$ = 6.0 $\times$ 10$^{21}$ cm$^{-2}$ based on the results from the spectral analysis of 
the Arc region (see Section \ref{subsec:arc}). Although the overall spectrum can be fitted by 
a $kT$ $\sim$ 9 keV thermal plasma, the fit is statistically rejected ($\chi^2/{\nu}$ $\sim$ 2.1) 
because of the large residuals in the soft band ($E$ $\sim$ 0.5--2 keV). When the LMC column is 
varied, the fit improves, but is still relatively poor ($\chi^2/{\nu}$ $\sim$ 1.5). The fitted 
$N_{\rm H,LMC}$ ($\sim$ 2.5 $\times$ 10$^{21}$ cm$^{-2}$) is significantly lower than that 
estimated for the Arc. Thus, the single plane-shock models with the LMC abundances are not 
appropriate to fit the observed spectrum. 

The fit improves to be statistically acceptable when the individual metal abundances are varied. 
We attempted to fit abundances for O, Ne, Mg, Si, S, Ar, Ca, and Fe while other elements (He, C, 
N, and Ni) were fixed at the LMC abundances. We found that only O and Ne abundances significantly
affected the fit. Thus, we varied O and Ne abundances and fixed other elements at the LMC values
($\chi^2/{\nu}$ = 64.8/55, Figure~\ref{fig:fig4}a and Table~\ref{tbl:tab2}). The best-fit electron 
temperature is $kT$ = 3.08$^{+0.63}_{-0.92}$ keV with a low ionization timescale of $n_{\rm e}t$ 
$\sim$ 0.60 $\times$ 10$^{10}$ cm$^{-3}$ s. We fixed the LMC column at $N_{\rm H,LMC}$ = 6.0 
$\times$ 10$^{21}$ cm$^{-2}$ in this fit. The best-fit abundances for O (= 0.03$^{+0.02}_{-0.01}$) 
and Ne (= 0.09$^{+0.04}_{-0.03}$) are low, showing no evidence for metal-rich ejecta. When the LMC 
column was varied, the fit does not statistically improve ($\chi^2/{\nu}$ = 62.8/54), and the 
best-fit $N_{\rm H,LMC}$ (= 1.2 $\times$ 10$^{22}$ cm$^{-2}$) is unreasonably high for the LMC. 
The best-fit abundances for O ($\sim$0.3) and Ne ($\sim$0.2) are similar to those of the LMC, but 
are poorly constrained. Thus, the E region spectrum can be described by a single plane shock model 
in which O and Ne abundances are lower than the LMC values.

Alternatively, we fit the spectrum with a two-component plane shock model with the metal abundances 
fixed at the LMC values for both components. The LMC column was fixed at $N_{\rm H,LMC}$ = 6.0 
$\times$ 10$^{21}$ cm$^{-2}$ for both components. The fit is statistically acceptable ($\chi^2/{\nu}$ 
= 60.7/54, Table~\ref{tbl:tab2}). The best-fit electron temperatures are $kT_{\rm soft}$ = 
1.72$^{+1.63}_{-0.35}$ and $kT_{\rm hard}$ = 3.20$^{+2.55}_{-1.06}$ keV. The soft component emission 
originates from an under-ionized thermal plasma ($n_{\rm e}t_{\rm soft}$ $\sim$ 3.4 $\times$ 10$^{10}$ 
cm$^{-3}$ s). The ionization timescale for the hard component is unconstrained. We also considered the 
hard tail of the observed X-ray spectrum to be nonthermal in origin, and thus repeated the two-component 
model fits by replacing the hard component plane shock model with a PL. The best-fit PL photon index 
is $\Gamma$ = 2.23$\pm$0.33, and the electron temperature of the thermal component is estimated to be 
$kT$ = 1.72$^{+2.38}_{-0.69}$ ($\chi^2/{\nu}$ = 56.2/55, Table~\ref{tbl:tab2}). The modeled PL spectrum 
is consistent with that for typical synchrotron radiation from the shock-accelerated electrons. In fact, 
synchrotron radiation observed in the Arc region shows a similar PL spectrum with $\Gamma$ = 2.4 
(Section~\ref{subsec:arc}).

\subsection{\label{subsec:south} Soft South Region}

The faint southern boundary region, which is distinctively red in Figure~\ref{fig:fig1}a, is spectrally 
soft. This has not been recognized by previous data, and our new {\it Chandra} data reveal this feature. 
We extracted $\sim$900 counts from this region (the ``S'' region in Figure~\ref{fig:fig2}). The observed 
X-ray spectrum shows line emission features at $E$ $\sim$ 0.6--1.5 keV as well as the hard tail at 
$E$ $>$ 3 keV (Figure~\ref{fig:fig4}b). We estimated the effects by the scattered photons from the 
pulsar and PWN due to the broad PSF wings of the mirror using the same methods described in 
Section~\ref{subsec:east}. The contamination by scattered photons from the pulsar and PWN is estimated 
to be $\sim$20\% in the 3-7 keV band in this region.  We fixed the PL component of the scattered 
photons (with the flux normalized for the area of the S region) in the spectral fits of the S region. 
Initially, we fit the spectrum with a single temperature NEI plane-shock model. We fixed the individual 
metal abundances at the LMC values, and also fixed the LMC column at $N_{\rm H,LMC}$ = 6.0 $\times$ 
10$^{21}$ cm$^{-2}$. The best-fit model ($kT$ $\sim$ 1.14 keV and $n_{\rm e}t$ $\sim$ 0.9 $\times$ 
10$^9$ cm$^{-3}$ s) is statistically poor ($\chi^2/{\nu}$ = 1.6). Varying $N_{\rm H,LMC}$ does not 
improve the fit ($N_{\rm H,LMC}$ $\sim$ 5.1 $\times$ 10$^{21}$ cm$^{-2}$, $\chi^2/{\nu}$ = 1.5). 
The poor fits are primarily caused by residuals at $E$ $\sim$ 0.7--0.9 keV. Because the residual line 
features at $E$ $\sim$ 0.7--0.9 keV appear to be the Fe L line complex, we varied the Fe abundance. 
The best-fit model ($kT$ $\sim$ 1.20$^{+0.19}_{-0.60}$ keV and $n_{\rm e}t$ = 0.33$^{+0.11}_{-0.07}$ 
$\times$ 10$^{10}$ cm$^{-3}$ s) statistically improves ($\chi^2/{\nu}$ = 32.9/33, Figure~\ref{fig:fig4}b 
and Table~\ref{tbl:tab2}), and suggests overabundant Fe (= 18.5$^{+59.5}_{-13.7}$). We similarly 
attempted to vary O and/or Ne abundances, and found that O and Ne overabundances are not required 
to improve the fit. When O and/or Ne abundances are fitted with the Fe abundance fixed at the LMC 
value, the best-fit model fit clearly shows residuals at E $\sim$ 0.7--0.9 keV, which suggests that 
the emission features in the 0.7-0.9 keV band are likely dominated by the broad Fe L line complex. 
Varying $N_{\rm H,LMC}$ results in nearly the same spectral parameters with no statistical improvement 
($\chi^2/{\nu}$ = 32.3/32): The Fe overabundance is still suggested (Fe = 17.3$^{+48.4}_{-13.1}$) 
with similar $kT$ ($\sim$1.37 keV) and $N_{\rm H,LMC}$ ($\sim$5.6 $\times$ 10$^{21}$ cm$^{-2}$). 

For completeness, we attempted two-temperature plane-shock models to fit the S region spectrum. 
$N_{\rm H,LMC}$ was fixed as above, and the metal abundances were also fixed at the LMC values. 
The best-fit model ($\chi^2/{\nu}$ = 31.4/31) suggests a high temperature plasma ($kT_{\rm hard}$ 
$>$ 2.4 keV) in addition to the soft component ($kT_{\rm soft}$ $\sim$ 0.35 keV). Because the single 
temperature model fit suggested an Fe overabundance, we tested whether the Fe overabundance is required 
in the two component model. When the Fe abundance is varied, an Fe overabundance ($>$1.2) is suggested, 
while other spectral parameters remain generally consistent with those estimated with the LMC abundances 
($\chi^2/{\nu}$ = 25.8/30). Nonetheless, these two-component models do not statistically improve the fit, 
compared to the single component models. Based on the single plane shock model fits, the enhanced Fe 
abundance appears to required to adequately fit the observed spectrum of the S region. However, the 
statistical uncertainties on the Fe abundance measurements are large (i.e., the Fe overabundance is a 
$\sim$2--3$\sigma$ detection above the LMC value). Also, the two-temperature plasma models may equally 
fit the observed spectrum without overabundant Fe. Thus, while the Fe overabundance is suggested in the 
S region, a firm conclusion may not be drawn with the current data.

\subsection{\label{subsec:west} Bright West Regions}

In contrast to the faint E and S regions, the western parts of the SNR show bright filamentary
emission features (Figure~\ref{fig:fig1}a). Hwang et al. (2001) revealed that the bright western 
filaments, except for the small region near the outer boundary in the direction opposite to the E 
arc, are dominated by soft X-ray emission from thermal plasma with $kT$ $\sim$ 0.5--1 keV. The 
previous {\it Chandra} data indicated no evidence for enhanced metal abundances in the bright western 
regions \citep{hwang01}, while the {\it XMM-Newton} RGS data suggested enhanced emission from Fe L 
and Ne K lines \citep{van01}. Results from our spectral analysis of the western filaments are generally 
consistent with those by Hwang et al. (2001). We here present our spectral analysis of two representative 
regions: the ``SW'' and ``NW '' regions (Figure~\ref{fig:fig2}). We choose the SW region because it 
shows bright thermal emission at the outer boundary, likely representing the blast wave with good photon 
statistics. The NW region includes a significant fraction of the hard emission ($kT$ $\sim$ 5 keV) among 
the western filaments (besides the regions of the candidate nonthermal filament at the outer boundary, 
Hwang et al. 2001). In the spectra of the SW and NW regions, the contamination by the scattered hard X-ray 
photons from the pulsar and PWN is negligible thanks to the small extraction area. Thus, we do not include 
the PL component of the broad PSF wings in the spectral fits of these regions.

\begin{figure}[]
\figurenum{4}
\centerline{\includegraphics[angle=0,width=0.9\textwidth]{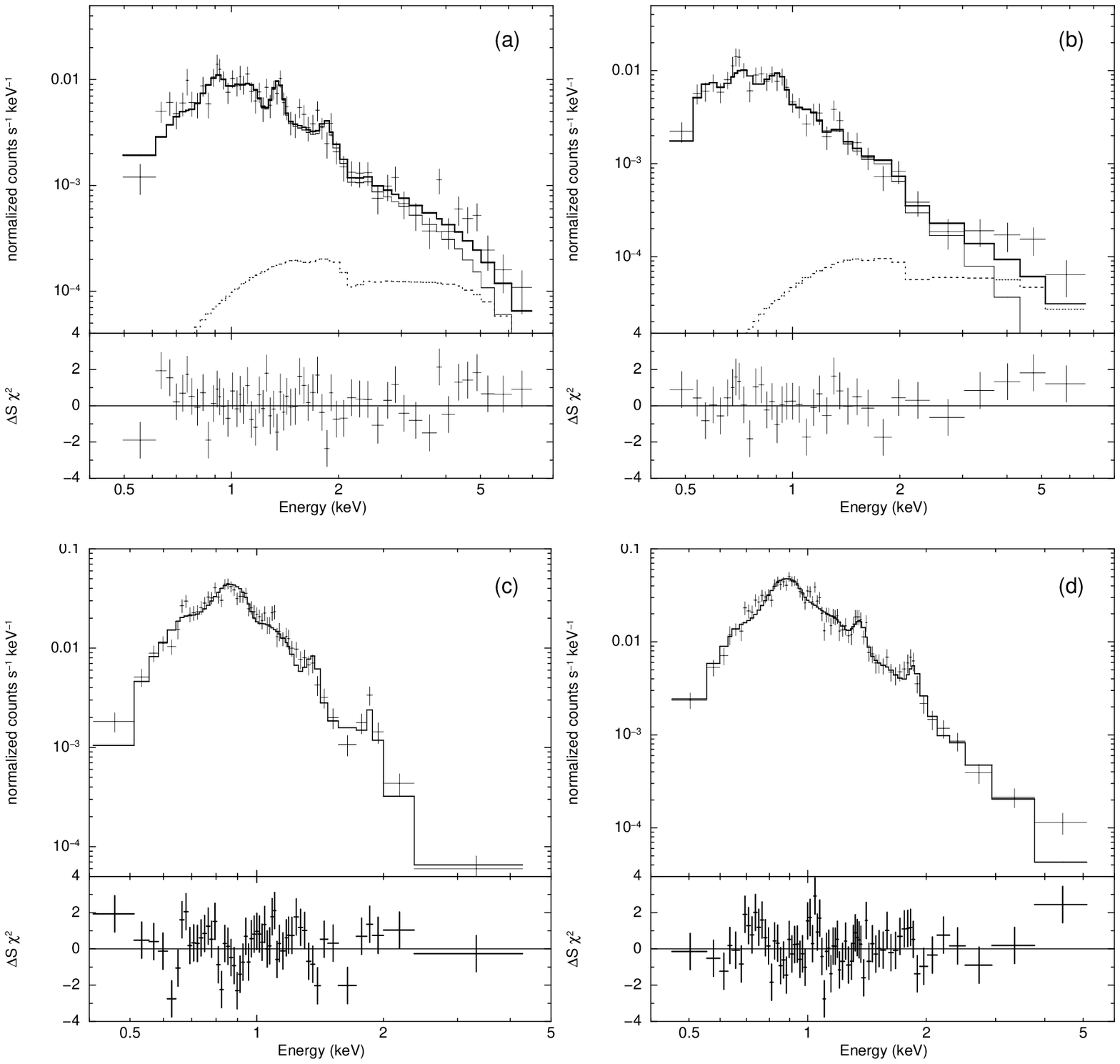}}
\figcaption[]{The {\it Chandra} ACIS spectrum of (a) the E, (b) the S, (c) the SW, and (d) 
the NW regions. In (a)-(d), the thick solid line shows the best-fit single plane-shock model. 
The lower panel shows the residuals from the best-fit model. In (a) and (b), the dotted
line shows a PL model representing the estimated contamination by the scattered photons
from the PWN.
\label{fig:fig4}}
\end{figure}

The SW region spectrum ($\sim$2200 counts) is presented in Figure~\ref{fig:fig4}c. The overall spectrum 
can be fitted by a single component plane-shock model ($kT$ $\sim$ 0.74 keV, $n_{\rm e}t$ $\sim$ 5.3 
$\times$ 10$^{10}$ cm$^{-3}$ s, $\chi^2/{\nu}$ = 74.5/54). We fixed the LMC column at $N_{\rm H,LMC}$ = 
6.0 $\times$ 10$^{21}$ cm$^{-2}$, and the individual metal abundances at the LMC values. When $N_{\rm 
H,LMC}$ was fitted, the best-fit value ($N_{\rm H,LMC}$ $\sim$ 6.9 $\times$ 10$^{21}$ cm$^{-2}$) is 
similar to that estimated from the Arc, and the overall fit does not improve ($\chi^2/{\nu}$ = 74.0/53). 
Varying the individual abundances results in the LMC-like abundances ($\sim$0.1--0.3) without a significant 
improvement in the fit ($\chi^2/{\nu}$ = 60.8/49) based on an F-test (F-probability $\sim$ 0.07). There 
is no evidence for the presence of a hard component ($E$ $\ga$ 3 keV) or metal overabundances in this region. 
The best-fit spectral parameters assuming the LMC abundances are summarized in Table~\ref{tbl:tab2}.

For the NW region spectrum ($\sim$2800 counts), a single plane-shock model fit with the fixed $N_{\rm 
H,LMC}$ and the LMC abundances is not acceptable ($\chi^2/{\nu}$ $\sim$ 2.0). When the LMC column was 
fitted, a $\sim$50\% higher column ($N_{\rm H,LMC}$ = 9.1$^{+1.5}_{-1.0}$ $\times$ 10$^{21}$ cm$^{-2}$, 
$kT$ $\sim$ 1 keV, $\chi^2/{\nu}$ $\sim$ 1.6) is implied. However, the observed spectrum of this
region is nearly identical to that of SW region in the soft energy band ($E$ $\la$ 1 keV, 
Figure~\ref{fig:fig4}), in which the large foreground absorption in the NW region, if it exists, should 
significantly affect the observed spectral shape. Then, we varied the metal abundances (with $N_{\rm 
H,LMC}$ = 6.0 $\times$ 10$^{21}$ cm$^{-2}$), which results in a significantly improved fit ($\chi^2/{\nu}$ 
= 80.1/65, Figure~\ref{fig:fig4}d). The best-fit abundances are low ($\sim$0.05--0.2) showing no evidence 
for metal-rich ejecta. The fitted spectral parameters are summarized in Table~\ref{tbl:tab2}. Since the 
hard spectral component with $kT$ $\sim$ 5 keV was suggested in this region in the previous work 
\citep{hwang01}, we, for completeness, attempted a two-component plane-shock model fit for the NW region. 
In this fit, we fixed $N_{\rm H,LMC}$ = 6.0 $\times$ 10$^{21}$ cm$^{-2}$ and the abundances at the LMC 
values. The best-fit model indicates $kT_{\rm soft}$ $\sim$ 0.6 keV and $kT_{\rm hard}$ $\sim$ 1.4 keV 
($\chi^2/{\nu}$ $\sim$ 1.0). Although the two temperature shock model can adequately fit the observed 
spectrum, the electron temperature for the hard component is significantly lower than that implied in 
the E region. Varying $N_{\rm H,LMC}$ and the abundances does not affect these results. The best-fit 
spectral parameters for the two component model are also summarized in Table~\ref{tbl:tab2}.

\section{\label{sec:disc} Discussion}

\subsection{\label{subsec:cr} Nonthermal Filament}

Using our deep {\it Chandra} observation, we firmly establish the nonthermal nature of the Arc in the E 
boundary of SNR 0540--69.3. The observed X-ray spectrum is well fitted by a PL with $\Gamma$ = 2.42, 
which is typical for synchrotron radiation from relativistically accelerated electrons in the strong shock. 
The contribution from underlying thermal emission is estimated to be small ($\sim$5\% of the total flux, 
if exists). Assuming the synchrotron origin, our SRCUT model fits indicate that the roll-off frequency 
between the radio and X-ray bands is $\nu_{\rm roll}$ $\sim$ 1--2 $\times$ 10$^{17}$ Hz. The {\it 
Chandra} image shows that the Arc consists of several substructures with various widths and brightness. 
Based on these filamentary features, we estimate the widths of the Arc to be in a range of 
$\sim$1$^{\prime\prime}$--2$\farcs$8 (FWHM) which correspond to a physical size of $\sim$0.3--0.7 pc. 
SNR 0540--69.3 may be in a free-expansion phase at the age of $\tau$ $\sim$ 1000 yr (e.g., Hwang et al.  
[2001] and references therein). Then, the angular distance of the Arc to the SNR center 
($\sim$35$^{\prime\prime}$ corresponding to a physical distance of $\sim$8.5 pc) suggests a 
time-average shock velocity $v_{\rm s}$ $\sim$ 8300 $\tau^{-1}_{1}$ km s$^{-1}$, where $\tau_{1}$ is 
the SNR age in units of 1000 yr. Assuming the Sedov phase, $v_{\rm s}$ $\sim$ 3300 $\tau^{-1}_{1000}$ 
km s$^{-1}$ is implied.  These velocities are in plausible agreement with the estimated shock 
velocities ($v_{\rm s}$ $\sim$ 2000--9000 km s$^{-1}$) for the SNR in a free-expansion or a Sedov 
phase, based on the spectral analysis of the thermal component emission for the E region 
(Section~\ref{subsec:therm}). The estimated widths of the Arc filaments and these shock velocities 
($v_{\rm s}$ $\sim$ 2000--9000 km s$^{-1}$ for the E region) imply the synchrotron loss time 
$\tau_{\rm loss}$ $\sim$ $\omega$~$v_s^{-1}$~$r$ $\sim$ 160--2700 yr, where $\omega$ is the 
advection distance of the downstream electrons or the physical width of the filament, and $r$ 
is the compression ratio in the shock. In this estimate, we assumed $r$ $\sim$ 5--8 for an 
efficient particle acceleration (e.g., Ellison et al. 2007). 

The peak frequency of a synchrotron emitting electron is $\nu_{\rm p}$ = 1.8 $\times$ 10$^{18}$
$E^2_{\rm e}$ $B_{\bot}$ Hz, where $B_{\bot}$ is the magnetic field component perpendicular to 
the velocity vector of the electron, and $E_{\rm e}$ is the electron energy. For $\nu_{\rm p}$ 
$\sim$ 3 $\times$ 10$^{17}$ Hz (or the photon energy $\sim$1.3 keV) representing typical X-ray 
photons as observed in the Arc spectrum, the corresponding electron energy $E_{\rm e}$ = 0.4 
$B_{\bot}$$^{-{1\over2}}$ erg implies $B_{\bot}$ = 136 $\tau_{\rm loss}$$^{-{2\over3}}$ G. 
For the estimated $\tau_{\rm loss}$ $\sim$ 160--2700 yr, we derive $B_{\bot}$ $\sim$ 7--46 $\mu$G.
Considering a geometrical projection effect in estimating the widths of the filaments, the actual $B$
field is likely larger than these estimates. For instance, the observed filament width would be 
overestimated by a factor of $\sim$5 assuming a spherical shock with an exponential emission profile 
\citep{bal06}. Then, $B$ $\sim$ 20--140 $\mu$G are estimated. The maximum electron energy is $E_{\rm 
e,max}$ = 2.5 $\times$ 10$^{-7}$ $\nu_{\rm roll}$$^{1\over2}$ $B_{\rm {\mu}G}$$^{-{1\over2}}$ TeV $\sim$ 
7--25 TeV. The estimated $B$ field is roughly an order of magnitude higher than typical LMC interstellar 
fields ($B$ $\sim$ 4 $\mu$G, Gaensler et al. 2005), indicating a strong magnetic field amplification 
in the SNR shock. It was suggested that the presence of ordered magnetic fields in the LMC (and in 
other young/irregular galaxies) could have originated from the CR-driven dynamo due to recent SN 
activity rather than the standard galactic rotational dynamo mechanism \citep{gaen05}. The detection 
of synchrotron emission by the shock-accelerated relativistic electrons in SNR 0540--69.3, which is 
located in the LMC's largest active star-forming region 30 Doradus, provides support for the presence 
of such a population of relativistic particles in the LMC. These spectral characteristics of the 
nonthermal emission, the morphology of the thin filamentary features in opposite sides of the SNR, 
and the young age of $\sim$1000 yr for SNR 0540--69.3 are similar to those observed in Galactic SNRs 
1006 (e.g., Bamba et al. 2003) and G330.2+1.0 \citep{park09}. While this general scenario is
plausible, we note that the quality of the current data is somewhat limited, and thus our
simplified interpretations of the magnetic field amplification need to be taken with a caution. 
For instance, the detailed spatial substructures of the Arc is complex rather than a spherical 
shell. Thus, our simple geometrical correction may not be fully justified, and the
true uncertainties on the magnetic field measurements could be significantly larger. High
angular resolution X-ray imaging data with deeper observations will be required to conclusively
reveal the true nature of the magnetic field structure in this SNR.

\subsection{\label{subsec:therm} Thermal Emission and Metal Abundances}

Based on the spectral analysis of the E region, we confirm the presence of a hard component ($E$ $>$ 
3 keV) in 0540--69.3. Two-component spectral model fits indicate that this hard component emission 
can be described either by a hot gas ($kT$ $\sim$ 3 keV) or a PL ($\Gamma$ $\sim$ 2.2). With the 
current data, it is difficult to statistically discriminate the thermal or nonthermal origin for the 
hard X-ray emission in the E region. Although the PL component is similar to that for synchrotron 
radiation in the Arc region ($\Gamma$ $\sim$ 2.4), there is no evidence for thin filamentary features 
which would have been anticipated along the outer boundary of the E region, had the hard X-ray emission 
originated from the shock accelerated relativistic electrons in highly amplified magnetic fields just 
behind the shock front. However, we note that SNR 0540--69.3 is significantly more distant than the 
Galactic SNRs that show thin nonthermal filaments. If the E region of 0540--69.3 consisted of a nest 
of thin filaments as observed in the Galactic nonthermal SNR G347.3--0.5, it could be seen largely as 
a faint smooth rim except for the brightest filaments in the Arc region. Thus, the nonthermal 
interpretations for the hard component emission in the E region are not clear with the current data. 
On the other hand, the hard tail of the E region spectrum can be described by a hot gas of $kT$ $\sim$ 
3 keV (either by a single or two component shock model fits). The presence of this hot plasma is plausible 
in young SNRs. Although a nonthermal origin for the hard X-ray emission in the E region cannot be ruled 
out with the current data, we here, for simplicity, discuss the nature of the E regional spectrum assuming 
a thermal origin. 

Based on the best-fit volume emission measure ($EM$ = $n_{\rm e}n_{\rm H}V$ = 1.74 $\times$ 10$^{57}$ 
cm$^{-3}$), estimated by the simple one-component plane-shock model fit, we derive the post-shock electron 
density $n_{\rm e}$ $\sim$ 1.44 $f^{-{1\over2}}$ cm$^{-3}$, where $f$ is the volume filling factor of 
the X-ray emission. In this estimate, we assumed $n_{\rm e}$ = 1.2$n_{\rm H}$ (where $n_{\rm H}$ is the 
H number density). We also assumed the X-ray-emitting volume $V$ $\sim$ 1.0 $\times$ 10$^{57}$ cm$^3$ for 
the E region with an {\it average} path-length of $\sim$2.4 pc along the line of sight, corresponding to 
the angular thickness of the E region ($\sim$10$^{\prime\prime}$) at the distance of $d$ = 50 kpc. The 
two-temperature model implies $n_{\rm e}$ $\sim$ 0.67 and 1.38 $f^{-{1\over2}}$ cm$^{-3}$ for the soft 
and the hard component, respectively. These estimated post-shock $n_{\rm e}$ are several times lower than 
those measured in the bright W regions (see below). The estimated ionization timescales ($n_{\rm e}t$ $\sim$ 
0.6--3 $\times$ 10$^{10}$ cm$^{-3}$ s) indicate the ionization ages $t$ $\sim$ 130--1600 yr depending on 
models. The swept-up mass in this region, $M_{\rm sw}$ = $n_{\rm H}$$m_{\rm p}$$V$ (where $m_{\rm p}$ is 
the proton mass), is estimated to be small ($M_{\rm sw}$ $\sim$ 1.0--1.4 $f^{1\over2}$ $M_{\odot}$), which 
is consistent with the shock propagating through a low-density medium in a young SNR. 

In contrast, the X-ray spectrum in the bright W regions can be fitted by thermal plasmas with relatively 
low electron temperatures ($kT$ $\sim$ 0.5--1.5 keV). These spectra also imply low LMC-like metal 
abundances. Thus, the bulk of X-ray emission in the W portions of the SNR originates from the shocked 
interstellar/circumstellar medium. Based on the fitted $EM$, we estimate the post-shock $n_{\rm e}$ for 
the SW and NW regions. In these calculations, we adopt the same method as above. For the SW region, 
we use the X-ray emitting volume $V$ $\sim$ 9 $\times$ 10$^{55}$ cm$^{-3}$, assuming the path-length of 
$\sim$1 pc along the line of sight. The estimated post-shock electron density is $n_{\rm e}$ $\sim$ 5.6 
$f^{-{1\over2}}$ cm$^{-3}$.  For the NW region, we assume $V$ $\sim$ 5 $\times$ 10$^{55}$ cm$^{-3}$ for 
the path-length of $\sim$0.8 pc along the line of sight. The estimated density is $n_{\rm e}$ $\sim$ 
6.8--12.3 $f^{-{1\over2}}$ cm$^{-3}$ depending on models. The blast wave is encountering several times 
denser ambient medium in western regions than in eastern regions. While the surrounding density is 
significantly different between E and W sides of the SNR, the overall radial extent is nearly symmetric 
in E-W as well as in N-S. We speculate that the ambient density is higher in W, and that the blast wave 
has probably entered the dense medium there recently. Although the H{\small I} map of the LMC \citep{kim98} 
shows that SNR 0540--69.3 is in fact surrounded by relatively bright H{\small I} filaments, these 
H{\small I} filaments are primarily located in N and E sides of the SNR. The bright W shell is not 
detected in the mid-IR \citep{will08}. On the other hand, Hwang et al. (2001) noted that the CO map of 
the LMC \citep{cohen88} shows molecular clouds to W of the SNR. However, the low-resolution of the CO 
map ($\sim$8$\farcm$8) does not allow one to determine whether the CO clouds are close enough to the SNR 
to account for the higher implied density in the western filaments. Alternatively, the dense material in 
the W regions of the SNR might be a relic structure of the stellar winds from the massive progenitor. 
In this scenario, however, the stellar winds must be highly asymmetric between E and W, which is difficult 
to reconcile with the winds from a 20--25 $M_{\odot}$ progenitor in the red supergiant phase. 
High-sensitivity, high-resolution radio observations of 0540--69.3 regions would be useful to study the 
structures of the surrounding medium and its interactions with the SNR, which may help reveal the origin 
of the bright W shell of the SNR.

If the SNR is in the adiabatic phase, we can derive the Sedov parameters of the SNR using the results 
of our spectral analysis. We attempt these estimates using the characteristic E region with a 
representative electron temperature of $kT$ = 3 keV. Assuming an electron-ion temperature equilibrium, 
this temperature implies a shock velocity $v_{\rm s}$ $\sim$ 1600 km s$^{-1}$. However, the assumed 
electron-ion temperature equilibrium is unlikely established for $v_{\rm s}$ $\ga$ several 10$^2$ 
km s$^{-1}$ \citep{ghav07}. Thus, $\sim$1600 km s$^{-1}$ may only be considered as a lower limit. 
The radial distance from the pulsar to the outer boundary in the E region is $\sim$30$^{\prime\prime}$ 
corresponding to $\sim$7.3 pc, and thus the Sedov age of $\tau_{\rm sed}$ $<$ 1800 yr are estimated 
for $v_{\rm s}$ $>$ 1600 km s$^{-1}$. This SNR age limit is in good agreement with the previously 
estimated SNR ages of $\sim$760--1660 yr. Alternatively, the SNR might still be in the free-expansion 
phase \citep{hwang01}, especially in the eastern parts where the density is low. Assuming the 
free-expansion phase, the shock velocity should be $v_{\rm s}$ $\sim$ 7150 $\tau^{-1}_{1}$ km s$^{-1}$. 
If we take $\tau_{\rm SNR}$ $\sim$ 800 and 1800 yr as {\it lower} and {\it upper limits} for the SNR 
age appropriate for the spectral parameters derived in the E region, the corresponding shock velocities 
are $v_{\rm s}$ $\sim$ 4000--9000 and 1600--3600 km s$^{-1}$ for a free-expansion and a Sedov phase, 
respectively. On the other hand, in the SW region, the electron temperature of $kT$ = 0.74 keV implies 
$v_{\rm s}$ $>$ 800 km s$^{-1}$. The radial distance from the SNR center to the SW boundary is also 
$\sim$30$^{\prime\prime}$, and thus $\tau_{\rm sed}$ $<$ 3600 yr is estimated. While this age range 
is broadly consistent with the previous estimates, the derived upper limit on the SNR age is larger 
than that estimated in the E region. Considering the fact that the angular extent is similar between 
the E and W boundaries, the large Sedov age estimated in western regions may support that the blast 
wave has recently entered a dense medium in western regions as discussed above.

The spectrally-soft emission in the S region appears to be caused by enhanced emission from the Fe L 
line complex (the one-shock model fits) and/or the low temperature of the plasma (the two-shock model 
fits). A low electron temperature ($kT$ $\sim$ 0.35 keV) for this region is implied by the two-component 
shock model fits. This temperature is $\sim$5--10 times lower than those estimated in the E region, 
which could suggest a significantly higher density and thus substantial brightening in the X-ray intensity 
in the S region compared with its surroundings, unless the low temperature component originates from 
small clumpy material that may not be resolved by the current data.  However, the overall X-ray emission 
in the S region is faint with no evidence for large density and high X-ray intensity, which is generally 
not consistent with the inferred low temperature. In the two-shock model fits, the existence of the hard 
component ($kT$ $>$ 2.4 keV) is not firmly conclusive due to the limited photon statistics (only $\sim$40 
source counts at $E$ $>$ 3 keV). In fact, the one- and two-shock model fits are statistically 
indistinguishable (Section~\ref{subsec:south}), and the presence of the two characteristic plasma 
components may not be convincingly justified in the current data. 

The single shock model fit suggests Fe overabundance in the S region. Although the detection of the 
Fe overabundance in the S region is not conclusively established in the current data, if true, the 
suggested candidate Fe-rich ejecta near the southern outer boundary of the core-collapse SNR is 
intriguing. The presence of the Fe overabundance in the S region would imply that the Fe-rich 
ejecta created at the deepest core of the massive progenitor has been expelled out to the outermost 
boundary of the SNR. Such a case has been detected in the young Galactic O-rich SNR Cas A (e.g., 
Hughes et al. 2000, Hwang \& Laming 2003). Although this general scenario is intriguing, a firm 
detection of the Fe ejecta in the S region is required before discussing conclusive and extensive 
interpretations of this candidate Fe ejecta. High photon statistics with deeper X-ray observations 
and improved NEI models to accurately account for the significantly under-ionized Fe L lines are 
required to perform a quantitative abundance study of this soft feature in the S region, and thus 
to reveal the true nature of this candidate Fe ejecta.

We do not find evidence for enhanced line emission from Ne and Fe in the bright W regions. This is 
inconsistent with the results reported by van der Heyden et al. (2001) using the {\it XMM-Newton} 
RGS data \citep{van01}, and the discrepancy remains unanswered. Van der Heyden et al. (2001) claimed 
moderate overabundances (by a factor of $\sim$2--3) in Ne and Fe relative to the O abundance in the 
W regions, without quoting uncertainties. We note that our {\it Chandra} data hint for less abundant 
O than Ne and Fe (e.g., in the bright NW region, we estimate the O abundance is $\sim$20\% of the 
LMC value, while the Fe abundance is $\sim$50\% of the LMC). Thus, we speculate that {\it high 
abundances} for Fe and Ne relative to O based on the XMM-Newton data could have been due to 
low-abundant O rather than the enhanced Ne and Fe. Finally, we find no evidence for the O-rich 
ejecta corresponding to the optical ring of [O {\small III}] enhancements close to the PWN. 
For the annular region of $R$ = 4$^{\prime\prime}$--5$^{\prime\prime}$ (the ``ring'' region in 
Figure~\ref{fig:fig2}), the observed X-ray spectrum can be fitted by a PL model with $\Gamma$ = 
2.39$\pm$0.09 ($N_{\rm H,LMC}$ = 6.0$\pm$0.5 $\times$ 10$^{21}$ cm$^{-2}$, $\chi^2_{\nu}$ = 
163.3/123). We place a 3$\sigma$ upper limit of $f$ $<$ 5.2 $\times$ 10$^{-16}$ erg cm$^{-2}$ 
s$^{-1}$ (= 5.0 $\times$ 10$^{-7}$ photons cm$^{-2}$ s$^{-1}$) for the O Ly$_{\alpha}$ line flux. 

\section{\label{sec:sum} Summary}

We performed a deep $\sim$120 ks {\it Chandra} observation of the O-rich SNR 0540--69.3 in the LMC. 
We conclusively establish the nonthermal nature of the ``arc'' in the E boundary of the SNR. The 
observed X-ray spectrum shows a PL continuum ($\Gamma$ $\sim$ 2.4), indicating synchrotron radiation 
from the shock-accelerated relativistic electrons. Thus, CR electron acceleration is evidently 
effective in the shock front of this young SNR. Assuming that synchrotron radiation is responsible 
for the observed X-ray and radio emission in this nonthermal filament, we estimate a roll-off 
frequency of $\nu_{\rm roll}$ $\sim$ 1--2 $\times$ 10$^{17}$ Hz. From our estimates of the shock 
velocities ($v_{\rm s}$ $\sim$ 2000--9000 km s$^{-1}$), the physical width of the filament, and 
the observed spectral shape, we derive post-shock magnetic fields $B$ $\sim$ 20--140 $\mu$G. The 
maximum electron energy is estimated to be $E_{\rm e,max}$ $\sim$ 7-25 TeV. 

We confirm the presence of the hard component emission in this young SNR primarily in the faint 
eastern regions. The hard component spectrum can be described either by a hot gas with $kT$ $\sim$ 
3 keV or a PL with $\Gamma$ $\sim$ 2.2. It is difficult to discriminate the origin of the hard 
component (thermal vs. nonthermal) because of the low photon statistics in the current data. 
If the hard X-ray emission is nonthermal, the estimated PL photon index suggests synchrotron 
radiation for its origin, which is similar to that detected in the Arc region. If the hard 
emission is thermal, the hot gas implies an ambient density of $n_0$ $\sim$ 0.17--0.35 cm$^{-2}$ 
which is several times lower than that estimated in the western regions ($n_0$ $\sim$ 1.4--3 
cm$^{-2}$). The W regions are dominated by bright soft filamentary emission features representing 
the shocked high density medium of lower electron temperatures ($kT$ $\sim$ 0.5-1.5 keV). Metal 
abundances in the W regions are low ($\la$ 0.3), indicating interstellar and/or circumstellar 
origins rather than metal-rich stellar ejecta. Our spectral analysis of thermal X-ray emission 
features indicates high shock velocities ($v_{\rm s}$ $\sim$ 2000--9000 km s$^{-1}$) which are 
usually expected in young SNRs where particle acceleration is efficient. 

In the S boundary of the SNR, we detect a faint emission feature which shows a distinctively soft 
X-ray spectrum compared to the surrounding regions. The Fe abundance in this region appears to be 
enhanced, although the detection of the Fe overabundance is not conclusive because of the low 
photon statistics in the currrent data and the incomplete Fe L line models for highly under-ionized
thermal plasma. If it is confirmed by follow-up high statistics data and improved NEI models, 
the suggested Fe-rich ejecta candidate at the outer boundary of the SNR is intriguing. It would be 
the detection of Fe ejecta created in the deepest core of the SN and subsequently expelled to 
nearly the outer blast wave just like those seen in Cas A. Deeper {\it Chandra} observations and
improved Fe L line models for significanly under-ionized plasma are required to reveal the true 
nature of this candidate Fe-rich ejecta and its implications on the core-collapse nucleosynthesis 
and the details of the explosion mechanism.

\acknowledgments

The authors thank the referee for the careful review that help improve this work.
We also thank K. Borkowski for providing us the updated atomic database and for the
extensive discussion on the Fe L line fits in the low-ionization plasma. S.P. also
thank S. Zhekov for the helpful discussion on the Fe L lines. S.P. and K.M. thank 
F. Bocchino for a useful discussion. This work was supported in part by Smithsonian 
Astrophysical Observatory under {\it Chandra} grant GO5-6063X.

\clearpage

\begin{deluxetable}{ccccccc}
\footnotesize
\tablecaption{Best-Fit Model Parameters for the Arc Region.
\label{tbl:tab1}}
\tablewidth{0pt}
\tablehead{ & \colhead{$N_{\rm H,LMC}$} & & \colhead{$\nu_{\rm roll}$} & 
\colhead{1 GHz Flux} & & \\
Model & \colhead{(10$^{21}$ cm$^{-2}$)} & \colhead{$\Gamma$} & \colhead{(10$^{17}$ Hz)} & 
\colhead{(mJy)} & $\alpha$ & $\chi^2$/$\nu$ }
\startdata
PL & 6.5$^{+1.8}_{-1.5}$ & 2.42$^{+0.26}_{-0.23}$ & - & - & - & 39.8/31 \\
SRCUT & 5.7$^{+1.0}_{-0.9}$ & - & 1.0$^{+1.4}_{-0.4}$ & 0.19$\pm$0.02 & 0.41\tablenotemark{a} 
& 41.6/31 \\
SRCUT & 5.7$^{+1.0}_{-0.8}$ & - & 2.1$^{+4.7}_{-1.2}$ & 7\tablenotemark{b} & 0.62$\pm$0.01 
& 41.1/31 \\
\enddata
\tablenotetext{a}{The radio spectral index $\alpha$ (where $S_\nu$ $\propto$ $\nu^{-\alpha}$)
is fixed at $\alpha$ = 0.41 as estimated for the entire SNR excluding the PWN \citep{man93b}.}
\tablenotetext{b}{The 1 GHz flux is fixed at $f_{\rm 1 GHz}$ = 7 mJy as we estimate based
on the 1.5 GHz image by Manchester et al. (1993).}

\end{deluxetable}

\begin{deluxetable}{ccccccccc}
\tabletypesize{\footnotesize}
\tablecaption{Best-Fit Spectral Parameters of SNR 0540--69.3.
\label{tbl:tab2}}
\tablewidth{0pt}
\tablehead{ & \colhead{$kT_{\rm soft}$} & \colhead{$kT_{\rm hard}$} & &
\colhead{$n_{\rm e}t_{\rm soft}$} & \colhead{$n_{\rm e}t_{\rm hard}$} & 
\colhead{$EM_{\rm soft}$} & \colhead{$EM_{\rm hard}$} & \\ 
\colhead{Region} & \colhead{(keV)} & \colhead{(keV)} & \colhead{$\Gamma$} &
\colhead{(10$^{10}$ cm$^{-3}$ s)} & \colhead{(10$^{10}$ cm$^{-3}$ s)} & 
\colhead{(10$^{57}$ cm$^{-3}$)} & \colhead{(10$^{57}$ cm$^{-3}$)} & 
\colhead{$\chi^2$/$\nu$} }
\startdata
E\tablenotemark{a} & 3.08$^{+0.63}_{-0.92}$ & - & - & 0.60$^{+0.12}_{-0.08}$ & - & 1.74$^{+0.47}_{-0.27}$ 
& - & 64.8/55\\
 & 1.72$^{+1.63}_{-0.35}$ & 3.20$^{+2.55}_{-1.06}$ & - & 3.37$^{+3.03}_{-1.47}$ & - & 0.37$^{+0.13}_{-0.08}$ 
& 1.59$^{+0.52}_{-0.40}$ & 60.7/54 \\
& 1.72$^{+2.38}_{-0.69}$ & - & 2.23$\pm$0.33 & 3.46$^{+6.16}_{-2.06}$ & - & 0.30$^{+0.16}_{-0.11}$ 
& - & 56.2/55 \\
S\tablenotemark{b} & 1.20$^{+0.19}_{-0.60}$ & - & - & 0.33$^{+0.11}_{-0.07}$ & - & 1.14$^{+0.40}_{-0.23}$ 
& - & 32.9/33 \\ 
SW & 0.74$^{+0.10}_{-0.09}$ & - & - & 5.27$^{+3.43}_{-1.67}$ & - & 2.32$^{+0.35}_{-0.26}$ &
- & 74.5/54 \\
NW\tablenotemark{c} & 0.90$^{+0.16}_{-0.20}$ & - & - & 6.30$^{+9.90}_{-2.90}$ & - & 
6.33$^{+3.96}_{-1.68}$ & - & 80.1/65 \\
 & 0.59$^{+0.04}_{-0.04}$ & 1.43$^{+0.75}_{-0.43}$ & - & 50.7$^{+34.1}_{-17.6}$ & $<$0.08 &
4.92$^{+0.48}_{-0.63}$ & 1.93$^{+1.70}_{-0.85}$ & 66.3/68 \\
\enddata
%\vspace{-5mm}
\tablecomments{$N_{\rm H,LMC}$ is fixed at 6.0 $\times$ 10$^{21}$ cm$^{-2}$.
Errors are with 90\% confidence. For each region, one- and two-component plane-shock
model fits are presented. The metal abundances are fixed at the LMC values, unless
noted otherwise.}
\tablenotetext{a}{The O and Ne abundances are varied in the one-shock model.}
\tablenotetext{b}{The Fe abundance is varied while other elemental abundances are fixed at 
the LMC values.}
\tablenotetext{c}{The metal abundances are varied in the one-shock model, while abundances are
fixed at the LMC values in the two-shock model.}

\end{deluxetable}

%\clearpage

\begin{thebibliography}{}

\bibitem[Anders \& Grevesse 1989]{anders89} Anders, E. \& Grevesse, N. 
1989, Geochim. Cosmochim. Acta, 53, 197

\bibitem[Ballet 2006]{bal06} Ballet, J. 2006, AdSpR, 37, 1902

\bibitem[Bamba et al. 2003]{bamba03} Bamba, A., Yamasaki, R., Ueno, M., \& Koyama, K. 2003,
\apj, 589, 827

\bibitem[Badenes et al. 2009]{bad06} Badenes, C., Borkowski, K. J., Hughes, J. P., Hwang, U.,
\& Bravo, E. 2006, \apj, 645, 1373 

\bibitem[Borkowski et al. 2001]{bor01} Borkowski, K. J., Lyerly, W. J., \&
Reynolds, S. P. 2001, \apj, 548, 820

\bibitem[Cohen et al. 1988]{cohen88} Cohen, R. S., Dame, T. M., Garay, G., Montani, J., 
Rubio, M., \& Thaddeus, P. 1988, \apj, 331, L95

\bibitem[Dickey \& Lockman 1990]{dl90} Dickey, J. M. \& Lockman, F. J. 1990,
\araa, 28, 215

\bibitem[Ellison et al. 2007]{ell07} Ellison, D. C., Patnaude, D. J., Slane, P. O., Blasi,
P., \& Gabici, S. 2007, \apj, 661, 879

%\bibitem[Fryer \& Kusenko 2006]{fryer06} Fryer, C. L. \& Kusenko, A. 2006, \apjs, 163, 335

\bibitem[Gaensler et al. 2005]{gaen05} Gaensler, B. M., Haverkorn, M., Staveley-Smith,
L., Dickey, J. M., McClure-Griffiths, N. M., Dickel, J. R., \& Wolleben, M. 2005,
Science, 307, 1610

\bibitem[Ghavamian et al. 2007]{ghav07} Ghavamian, P., Laming, J. M., \& Rakowski, C. E.,
2007, \apj, 654, L69

\bibitem[Gotthelf \& Wang 2000]{gw00} Gotthelf, E. V. \& Wang, Q. D. 2000, \apj, 
532, L117

\bibitem[Green 2009]{green09} Green, D. A. 2009, A Catalog of Galactic Supernova Remnants
(2009 March version; Cambridge: Mullard Radio Astronomy Obs., Cavendish Laboratory),
http://www.mrao.cam.ac.uk/surveys/snrs

\bibitem[Hirayama et al. 2002]{hira02} Hirayama, M., Nagase, F., Endo, T., Kawai, N.,
\& Itoh, M. 2002, \mnras, 333, 603

\bibitem[Hughes et al. 2000]{hughes00} Hughes, J. P., Rakowski, C. E., Burrows, D. N., \&
Slane, P. O. 2000, \apj, 528, L109

\bibitem[Hughes 2001]{hughes01} Hughes, J. P. 2001, in ``Young Supernova Remnants'',
The Eleventh Astrophysics Conference (Melville, NY: AIP), eds. S. S. Holt \& U. Hwang,
419

\bibitem[Hwang et al. 2001]{hwang01} Hwang, U., Petre, R., Holt, S. S., \&
Szymkowiak, A. E. 2001, \apj, 560, 742

\bibitem[Hwang \& Laming 2003]{hwang03} Hwang, U. \& Laming, J. M. 2003, \apj, 597, 362

\bibitem[Kaaret et al. 2001]{kaa01} Kaaret, P. et al. 2001, \apj, 546, 1159

\bibitem[Kim et al. 1998]{kim98} Kim, S., Staveley-Smith, L., Dopita, M. A., Freeman, 
K. C., Sault, R. J., Kesteven, M. J., McConnell, D. 1998, \apj, 503, 674

\bibitem[Kirshner et al. 1989]{kir89} Kirshner, R. P., Morse, J. A., Winkler, P. F.,
\& Blair, W. P. 1989, \apj, 342, 260

\bibitem[Manchester et al. 1993a]{man93a} Manchester, R. N., Mar, D. P., Lyne, A. G.,
Kaspi, V. M., \& Johnston, S. 1993a, \apj, 403, L29

\bibitem[Manchester et al. 1993b]{man93b} Manchester, R. N., Staveley-Smith, L., \&
Kesteven, M. J. 1993b, \apj, 411, 756

\bibitem[Mathewson et al. 1980]{math80} Mathewson, D. S., Dopita, M. A., Tuohy, I. R.,
\& Ford, V. L. 1980, \apj, 242, L73 

\bibitem[Middleditch \& Pennypacker 1985]{mp85} Middleditch, J. \& Pennypacker, C. R. 
1985, \nat, 313, 659

\bibitem[Morse et al. 2006]{morse06} Morse, J. A., Smith, N., Blair, W. P., Kirshner,
R. P., Winkler, P. F., \& Hughes, J. P. 2006, \apj, 644, 188

%\bibitem[Nomoto et al. 1997a]{nomoto97a} Nomoto, K., Hashimoto, M., Tsujimoto, T., Thielemann, 
%F.-K., Kishimoto, N., Kubo, Y., \& Nakasato, N. 1997a, Nuclear Physics A, 616, 79

%\bibitem[Nomoto et al. 1997b]{nomoto97b} Nomoto, K., Iwamoto, K., Nakasato, N., Thielemann, 
%F.-K., Brachwitz, F., Tsujimoto, T., Kubo, Y., \& Kishimoto, N. 1997b, Nuclear Physics A, 621,
%467

\bibitem[Park et al. 2009]{park09} Park, S., Kargaltsev, O., Pavlov, G. G., Mori, K.,
Slane, P. O., Hughes, J. P., Burrows, D. N., \& Garmire, G. P. 2009, \apj, 695, 431

\bibitem[Petre et al. 2007]{petre07} Petre, R., Hwang, U., Holt, S. S., Safi-Harb, S., 
\& Williams, R. M. 2007, \apj, 662, 988

\bibitem[Reynolds 1985]{rey85} Reynolds, S. P. 1985, \apj, 291, 152

\bibitem[Reynolds 1998]{rey98} Reynolds, S. P. 1998, \apj, 493, 375

\bibitem[Reynolds \& Keohane 1999]{rey99} Reynolds, S. P. \& Keohane, J. W. 1999, \apj,
525, 368

\bibitem[Russell \& Dopita 1992]{rd92} Russell, S. C. \& Dopita, M. A. 1992, \apj, 384, 508

\bibitem[Serafimovich et al. 2004]{sera04} Serafimovich, N. I., Shibanov, Yu. A., Lundqvist,
P., \& Sollerman, J. 2004, \aap, 425, 1041

\bibitem[Seward et al. 1984]{sew84} Seward, F. D., Harnden, F. R., Jr., \& Helfand, D. J.
1984, \apj, 287L, 19

\bibitem[Seward \& Harnden 1994]{sew94} Seward, F. D. \& Harnden, F. R. 1994, \apj, 421, 581

\bibitem[Smith et al. 2001]{smith01} Smith, R. K., Brickhouse, N. S., Liedahl, D. A., \&
Raymond, J. C. 2001, \apj, 556, L91

\bibitem[Townsley et al. 2000]{town00} Townsley, L. K., Broos, P. S., Garmire, 
G. P., \& Nousek, J. A. 2000, \apj, 534, L139
%\bibitem[Townsley et al. 2000]{town00} Townsley, L. K. et al. 2000, \apj, 534, L139

\bibitem[Townsley et al. 2002]{town02} Townsley, L. K., Broos, P. S., Chartas, 
G., Moskalenko, E., Nousek, J. A., \& Pavlov, G. G. 2002, Nucl. Instrum. Methods 
Phys. Res. A, 486, 716

\bibitem[van der Heyden et al. 2001]{van01} van der Heyden, K.J., Paerels, F., 
Cottam, J., Kaastra, J. S., \& Branduardi-Raymont, G. 2001, \aap, 365, L254

\bibitem[Williams et al. 2008]{will08} Williams, B. J. et al. 2008, \apj, 687, 1054

\end{thebibliography}
\end{document}